\documentclass[11pt,a4paper]{article}
\usepackage{amsmath,amssymb}
\usepackage{epsfig,graphicx}

\begin{document}

\title{\bf Quasi-classical Hawking Temperatures and Black Hole Thermodynamics}
\author{Terry~Pilling\footnote{{\bf e-mail}: Terry.Pilling@ndsu.edu}
\\
\small{\em Department of Physics, North Dakota State University} \\
\small{\em Fargo, North Dakota, USA, 58105}
}
\date{}
\maketitle

\begin{abstract}
The semi-classical derivation of Hawking radiation 
for axially symmetric, stationary spacetimes with a Killing horizon
is examined following the recent quasi-classical tunneling
analysis \cite{singleton2008} and a simple formula is found for 
the inverse Hawking temperature $\beta = 1/T_H$. 
The formula is invariant under canonical transformations and 
is shown to be equivalent to the integral of a closed differential
1-form around the horizon enclosing a pole. The Hawking temperature
is then given in terms of the winding number in the first homotopy 
group of the torus formed from the compactified imaginary parts of the
analytically continued radial and time variables.
\end{abstract}

\section{Introduction}

The semi-classical WKB formula \cite{padmanabhan1999,parikh2000} 
for deriving the Hawking temperature in spacetime backgrounds with 
a Killing horizon has attracted much attention in the recent literature.
The original analysis was motivated by the idea that the particles 
constituting the Hawking radiation should arise via tunneling out of 
the black hole.
Thus, one should be able to explicitly calculate the temperature 
in this way. 
This semi-classical `tunneling' formula has evolved significantly
since its inception due to several problems arising
when generalizing it to various black hole backgrounds and
alternate coordinate systems.
We will begin by describing these problems and their resolutions.

The first problem with the semi-classical tunneling
method is that it is not actually tunneling in the usual sense. 
Instead it is closer to an over-the-barrier reflection 
problem \cite{landau1958} since the particle momentum is never imaginary. 
However, there does appear an imaginary contribution to the tunneling 
amplitude due to the fact that the classical path of the particle 
has a pole at the horizon and so it is analytically continued around the pole. 

The next problem is that, although the original semi-classical method \cite{parikh2000}
works in the case of the Schwarzschild black hole in 
Painlev{\' e} coordinates, it gives incorrect temperatures in many 
other backgrounds and coordinates \cite{akhmedov2007}. 

Finally, the original formula is not canonically invariant as it must be in order that
it describe proper quantum mechanical observables \cite{chowdhury2008}.
We showed \cite{akhmedov2007} that this last problem can be corrected 
by using a canonically invariant formula, but with the result that all temperatures, 
including that found in Painlev{\' e} coordinates, then differed by a factor of two 
from the standard Hawking temperature. This was the so-called factor two problem.
Recently this factor two problem has been solved. The solution came via the discovery 
of a previously overlooked contribution from the transformation of 
the {\it time variable} \cite{singleton2008} coming from the fact that the time 
coordinate changes sign across the horizon. 
This new temporal contribution must be incorporated in order to get the correct 
Hawking temperatures. 
Although this new contribution fixes the factor of two problem, it also destroys the 
close analogy to semi-classical tunneling which motivated the tunneling 
analysis in the first place. In other words, {\it it is not the usual quantum 
mechanical tunneling that produces Hawking radiation}, the difference comes 
from the fact that general relativity treats time in a different way than conventional
quantum mechanics. We refer to the semi-classical method, including the 
new temporal contribution, as quasi-classical tunneling.

In the following, we begin with a summary of the semi-classical derivation
of the Hawking temperature along with the new contribution coming
from the time variable. Next, these contributions are shown to combine
into the integral of a differential 1-form around the horizon
which is therefore a canonically invariant result. However, the 
path is not closed in the time variable. We remedy this by changing
coordinates to compactify the imaginary part of the time coordinate 
at the horizon allowing us to derive the hawking temperature in terms 
of the integral of the new canonical momentum 1-form around the horizon.
We then use the 1-form to show that the temperature can be found in terms
of the periodicity in the Euclidean time as you circle the horizon. 
Finally a connection is made with black hole thermodynamics using a 
method similar to that given in \cite{pilling2008} wherein the fluctuations 
in the black hole mass lead to a similar formula. 

\section{Quasi-classical tunneling}

In this section we review the details of the canonically invariant
semi-classical method. The purpose is two-fold. First, many researchers
continue to use the original, incorrect, semi-classical formula and so it is 
deemed necessary to show explicitly why the canonically invariant one must be
used. Second, the resulting formulas lead, in the following section,
to the formulation of quasi-classical tunneling which is the subject of 
this work.

Consider an axially symmetric, stationary background with 
Killing horizon at $r = r_H$ where we choose the `mostly plus' metric signature
and natural units so that all of the constants are 1. 
A scalar field moving in this background is given, in the semi-classical 
approximation, by $\phi = e^{-\frac{iS}{\hbar}}$ where $S$ is the action. 
The Klein-Gordon equation
\begin{equation}
- \frac{\hbar^2}{\sqrt{-g}} \partial_\mu 
\left( \sqrt{-g} g^{\mu \nu} \partial_\nu \phi \right) + m^2 \phi = 0
\end{equation}
then reduces, at zeroth order in $\hbar$, to the Hamilton-Jacobi 
equation
\begin{equation}
g^{\mu \nu} \partial_\mu S \partial_\nu S + m^2 = 0,
\end{equation}
where $g_{\mu \nu}$ is an arbitrary 2-dimensional metric for the
$(t,r)$ section of the spacetime at the equator $\theta = \pi/2$. 

We now split the action into a temporal and spatial part,
$S = Et + S_0(\vec{x})$, and substitute into the Hamilton-Jacobi 
equation. The solution is
\begin{equation}
\label{partialS0}
\partial_r S_0 = \frac{1}{g_{00}} \left[
g_{01} E \pm \sqrt{-g} \sqrt{E^2 + g_{00} m^2} \right]
\end{equation}
which will be used below to find the black hole decay rate.

Let us now examine the lagrangian coming from the metric:
\begin{equation}
\label{lagrangian}
{\cal L} = \frac{1}{2} g_{\mu \nu} \frac{d x^\mu}{d \tau} \frac{d x^\nu}{d \tau},
\end{equation}
where $\tau$ is some affine parameter along the geodesic which
may be the proper time in the case of massive particles and
the usual affine parameter in the case of null geodesics.
For massless particles ${\cal L} = 0$ and for particles of 
mass $m$, $2{\cal L} = -m^2$.

This lagrangian defines the canonical momenta in the usual way as
\begin{equation}
\label{momenta}
\begin{split}
p_r &\equiv \frac{\partial {\cal L}}{\partial \dot{r}} 
= g_{01} \dot{t} + g_{11} \dot{r} \\
p_t &\equiv -\frac{\partial {\cal L}}{\partial \dot{t}} 
= - g_{00} \dot{t} - g_{01} \dot{r} \equiv E,
\end{split}
\end{equation}
where the dot denotes the derivative with respect to the affine
parameter $\tau$.
We can use $p_t$ to eliminate $\dot{t}$ from (\ref{lagrangian})
leaving an expression for the coordinate velocity in the 
radial direction
\begin{equation}
\label{dotr}
\dot{r} = \pm \sqrt{\frac{g_{00} m^2 + E^2}{-g}}
\end{equation}
where $+$ is for out-going particles and $-$ for in-going and
we have denoted the determinant of the metric by $g$.
Thus, the momentum component $p_r$ can be written as
\begin{equation}
\label{momentum}
p_r = - \frac{1}{g_{00}} \left[
g_{01} E \pm \sqrt{-g} \sqrt{E^2 + g_{00} m^2} \right].
\end{equation}
Comparing this expression with (\ref{partialS0}) we see that
$\partial_r S_0 = - p_r$, and we can solve for the spatial part 
of the action
\begin{equation}
\Delta S_0 = - \int p_r dr
\end{equation}
The decay rate is given by
\begin{equation}
\label{path}
\begin{split}
\Gamma &= \left| \left<x_f \right| \left. x_i \right> \right|^2 
= \left<x_i \right| \left. x_f \right> \left<x_f \right| \left. x_i \right> \\
&\sim e^{- \text{Im } \Delta S} = e^{\text{Im } \left(E \Delta t - \oint p_r dr \right)}
\end{split}
\end{equation}
where the canonically invariant integral is over the closed path as 
shown in \cite{akhmedov2007}.

In \cite{singleton2008} it was shown using Kruskal-Szekeres
coordinates that the time variable changes as
$t \rightarrow t-2 \pi i M$ as one crosses the horizon and
so, for the closed path in (\ref{path}), we will have 
$\Delta t = - 4 \pi i M$.
Thus the decay rate (\ref{path}) is
\begin{equation}
\Gamma \sim e^{-4 \pi M E} e^{- \text{ Im } \oint p_r dr} \equiv e^{-\beta E}.
\end{equation}
The inverse temperature $\beta = 1/T_H$ is then given by
\begin{equation}
\label{form1}
\beta E = 4 \pi M E + \text{ Im } \oint p_r dr.
\end{equation}
Now insert the expression for the canonical momentum (\ref{momentum})
to find
\begin{equation}
\beta = 4 \pi M - \frac{1}{E} \text{ Im } \oint \frac{
g_{01} E \pm \sqrt{-g} \sqrt{E^2 + g_{00} m^2}}{g_{00}} dr.
\end{equation}
One can see that there is a pole at the horizon $r = r_H$ since 
$g_{00}(r_H) = 0$ and so we shift the pole into the upper half plane 
$r_H \rightarrow r_H + i \epsilon$ and write 
$g_{00} = g^\prime_{00}(r_H) (r - r_H - i\epsilon)$
for the out-going part of the path and into the lower
half plane for the in-going part of the path. The result is
a positively oriented (counter-clockwise) path around the
horizon. The result is
\begin{equation}
\beta = 4 \pi M - \underset{\epsilon \rightarrow 0}{\text{lim }}
\text{Im } \oint \frac{
g_{01} \pm \sqrt{-g}}{g^\prime_{00}(r_H) (r - r_H \pm i\epsilon)} dr
\end{equation}
where we have set $\epsilon = 0$ in the mass term and 
the upper/lower signs correspond to out-going/in-going particles
respectively.

Let us examine this expression for $\beta$ in detail in order to
show how it corrects the differences that were encountered between
Painlev{\' e} and other coordinates systems in the original (incorrect)
semi-classical method.
We split the integral into the out-going and in-going parts
\begin{equation}
\label{formula1}
\begin{split}
\beta = 4 \pi M 
&- \frac{1}{g^\prime_{00}(r_H)} \underset{\epsilon \rightarrow 0}{\text{lim }}\\
&\text{Im } \left\{ 
\int_{r_i}^{r_f} \frac{g_{01} + \sqrt{-g}}{r - r_H - i\epsilon}
+ \int_{r_f}^{r_i} \frac{g_{01} - \sqrt{-g}}{r - r_H + i\epsilon}
\right\} dr
\end{split}
\end{equation}
where $r_i < r_H$ and $r_f > r_H$.
We see explicitly by this expression that, for metrics with an off-diagonal 
component $g_{01}$, there is
a difference in sign between the in-going term and the out-going term. 
This causes the $g_{00}$ terms to cancel out between the out-going and 
in-going terms and is precisely the reason why the amplitude is different 
for the in-going and out-going paths in coordinates, such as Painlev{\' e}, 
where $g_{01} \neq 0$. Thus if, as is often done, one simply squares the
amplitude coming from the out-going path, then the $g_{01}$ component will
not be canceled as it should be and the result is a factor of 2 multiplying $\beta$.
This can be seen by noticing that on the horizon we have $g_{01} = 1$ and 
$\sqrt{-g} = 1$ in Painlev{\' e} coordinates. 
Thus, although the original method gives the
correct Hawking temperature, this is a coincidence and only works in 
special coordinate systems. 
Continuing one step further, we can use 
\begin{equation}
\underset{\epsilon \rightarrow 0}{\text{lim }} \text{Im } \frac{1}{r - r_H \pm i \epsilon}
= \pi \delta(r - r_H)
\end{equation}
to get a simple expression for the inverse temperature
\begin{equation}
\beta = 4 \pi M - \left. \frac{2 \pi \sqrt{-g}}{g^\prime_{00}} \right|_{r = r_H}
\end{equation}

Later on, we will suggest a connection between quasi-classical tunneling and
black hole thermodynamics. To make this connection more apparent let us here take 
note of an alternate way of writing the inverse temperature given in (\ref{formula1}).
That is
\begin{equation}
\label{formula2}
\begin{split}
\beta &= 4 \pi M - 2 \text{ Im } \int_{r_i}^{r_f} \frac{\sqrt{-g}}{g_{00}} dr \\
&= 4 \pi M + i \oint \frac{\sqrt{-g}}{g_{00}} dr.
\end{split}
\end{equation}
The inverse temperature written in this way will appear when we examine the
first law of back hole thermodynamics using the method 
of \cite{pilling2008} in the penultimate section below. 

Let us now return to (\ref{form1}) and parameterize the path in terms of 
a parameter $\theta$ as follows
\begin{equation}
\label{form2}
\begin{split}
\beta E &= -\text{Im }E \Delta t + \text{Im } \oint p_r \frac{dr}{d\theta} d\theta \\
&= -\text{Im }\oint E \frac{dt}{d\theta} d\theta + \text{Im } \oint p_r \frac{dr}{d\theta} d\theta \\
&= \text{Im }\oint \left( -p_t \frac{dt}{d\theta} + p_r \frac{dr}{d\theta} \right) d\theta \\
&= \text{Im }\oint P_\mu \frac{dx^\mu}{d\theta} d\theta \\
&= \text{Im } \oint P_\mu dx^\mu
\end{split}
\end{equation}
where we have defined $P_0 = - p_t$ and $P_1 = p_r$.
We have abused the notation in the above formula since 
the path in the $t$ variable is not closed.
Both $r$ and $t$ are continued to complex values and both depend on the 
path via a parameter $\theta$.
Explicitly, we parameterize $r$ as $r = 2M + \epsilon e^{i \theta}$
as in \cite{akhmedov2007} then
\begin{equation}
t = \text{Re}(t) - 2Mi\theta
\end{equation}
so that 
\begin{equation}
\frac{dt}{d\theta} d\theta = -2Mi d\theta.
\end{equation}
When $\theta$ traverses the closed path, $\theta \rightarrow \theta + 2 \pi$, 
the $r$ variable returns to its original value, but 
the time coordinate changes by $t \rightarrow t - 4 \pi i M$ and
we see that the path is not closed in $t$.
Although this is not a problem {\it per se} in that our formula 
gives the correct temperatures, we can close the path in the imaginary 
part of the time variable and thereby give a more precise topological 
reason for our result.

The coordinate time diverges at the horizon, 
which is the reason for the sign change in $t$ as you cross the horizon 
and is the source of the temporal contribution to the tunneling amplitude.
We have seen this contribution by transforming to Kruskal coordinates
in \cite{singleton2008}.
Transforming the (complexified) $t$ variable\footnote{This transformation
is completely unnecessary for our result given in (\ref{formula2}). 
It is only done to close the path in both variables and make a connection 
to the homotopy group of the torus.} as
\begin{equation}
\label{timecoord}
T = e^{-t} = e^{-\text{Re}(t)} e^{2Mi\theta}
\equiv T_0 e^{2Mi\theta}
\end{equation}
has the effect of compactifying the imaginary part of the 
path in $t$. We then have have a closed path in the imaginary
parts of both the $t$ and the $r$ variable as one traverses 
the horizon. In fact we can write
\begin{equation}
E dt = E \frac{dt}{dT} dT = - \frac{E}{T} dT
\end{equation}
so that 
\begin{equation}
\label{form3}
\begin{split}
\beta E &= \text{Im } \oint P_\mu dx^\mu \\
&= \text{Im } \oint \left[ \frac{E}{T} dT + p_r dr \right] \\
&= \text{Im } \int_0^{2 \pi} \left[ \frac{E}{T} \frac{dT}{d \theta} 
+ p_r \frac{dr}{d \theta} \right] d \theta \\
&= \text{Im } \int_0^{2 \pi} \left[ 2 M i E + p_r \frac{dr}{d \theta} \right] d \theta \\
&= \text{Im } \left[ 4 \pi M i E +  4 \pi M i E \right]
\end{split}
\end{equation}
where we used the residue of the pole in $p_r$ to get the second term.
Thus $\beta = 8 \pi M$ as expected.
Notice that the form $P = P_\mu dx^\mu$ as given in (\ref{form3}) is
closed in complex $t-r$ space, but not exact (since there are poles). 
Thus it is an element of
the first homotopy group of the torus $\pi_1 ( S^1 \times S^1 )$ and
corresponds to traversing each of the generating loops once, giving the
same contribution to the temperature from each one as we have shown 
in (\ref{form3}) above.
One might expect that since we have formulated the expression for the temperature
in terms of a 1-form, the result should be coordinate invariant.
However, this is not so. The path in (\ref{form3}) is only closed
in complex $r$ and $t$ space; the real part of the path is not closed
by itself. This means that the Hawking temperature is not invariant under 
general coordinate transformations since any coordinate transformation 
which removes the pole on the horizon will result in a vanishing residue.

\section{Euclidean time}

We can use these results to make a connection to the Euclidean time formalism of
Gibbons, Hartle and Hawking \cite{hartle1976,gibbons1977} by eliminating the
$r$ dependence of the path in terms of $t$ alone.
Restrict consideration to massless particles and let the parameter $\theta$ 
introduced in the previous section be chosen as the affine parameter so that
\begin{equation}
\label{euclidean2}
\begin{split}
\beta E &= - \text{Im }\oint p_t \frac{dt}{d\tau} d\tau 
+ \text{Im }\oint p_r \frac{dr}{d\tau} d\tau \\
&= \text{Im }\oint \left(-g_{00} \dot{t}^2 + g_{11} \dot{r}^2 \right) d\tau
\end{split}
\end{equation}
The lagrangian (\ref{lagrangian}) can be used to eliminate the $\dot{r}^2$ term
\begin{equation}
\label{euclidean3}
\begin{split}
\beta E &= \text{Im }\oint \left(- g_{00} \dot{t}^2 - g_{00} \dot{t}^2 - 2 g_{01} \dot{r} \dot{t} \right) d\tau \\
&= 2 E \text{ Im }\oint \frac{dt}{d\tau} d\tau = 2 E \text{ Im } \Delta t \\
\Rightarrow \beta &= 8 \pi M
\end{split}
\end{equation}
This says that twice the imaginary part of the change in $t$ as you traverse 
the loop gives the Hawking temperature. 
As can be seen by our derivation of $\Delta t$ in Kruskal-Szekeres
coordinates, this is related to the result of Gibbons and Hawking \cite{gibbons1977}
which says that the period of the Euclidean time as you traverse a 
loop about the horizon is the Hawking temperature. 

\section{Blackhole thermodynamics}

A simple argument can be given to create a connection between our
quasi-classical formula and the first law of blackhole thermodynamics.
Black hole thermodynamics is a set of relations that are satisfied
by black holes and are usually called thermodynamical because
of their similarity to the usual laws of classical thermodynamics.
The temperature is defined via the first law of black hole 
thermodynamics as
\begin{equation}
dS = \frac{dM}{T} \equiv \beta \; dM
\end{equation}
where we are assuming a non-charged, non-rotating black hole.
The entropy of the black hole is defined in terms of the 
horizon radius as $S = A/4 = \pi r^2|_{r = r_H}$ and so the 
first law can be written as
\begin{equation}
\label{dSdM}
\beta = \frac{dS}{dM} = 2 \pi r \frac{d r_H}{dM}
\end{equation}
We will now assume a fluctuating mass (and therefore radius) and 
compute the resulting temperature \cite{pilling2008,zhang2008}.
In particular, the change in the mass could be caused by out-falling
matter, i.e. particle emission.
The result will be a formula for the temperature which bears a
close resemblance to our quasi-classical formula (\ref{formula2}) above. 

To see this connection explicitly consider again 
a counterclockwise oriented complex path surrounding the horizon radius 
and write $r_H$ as a residue integral over this path.
Since the Killing horizon is defined as the point where
$g_{00} = 0$ we will write this, for Schwarzschild coordinates, as
\begin{equation}
r_H = \frac{i}{2\pi} \oint \frac{\sqrt{-g}}{g_{00}} dr
\end{equation}
so that (\ref{dSdM}) becomes
\begin{equation}
\label{generalFormula}
\beta = i \frac{d r_H}{dM} \oint \frac{\sqrt{-g}}{g_{00}} dr
\end{equation}
For a Schwarzschild black hole $r_H = 2 M$ and so this immediately gives
\begin{equation}
\beta = 2 i \oint \frac{\sqrt{-g}}{g_{00}} dr
\end{equation}
which is just twice the spatial part of (\ref{formula2}).
Since the spatial and temporal parts contribute equally this
gives the same result as that of our quasi-classical analysis.

\section{Examples and potentially problematic cases}

The formula (\ref{generalFormula}) seems to be problematic
when the metric determinant is zero, or when the horizon
radius is at $r_H = 0$.
In this section we present some explicit examples 
where these problems occur and present resolutions
to them. We also apply the method to the cases of charged 
and rotating black holes.

{\bf Schwarzschild metric in isotropic coordinates:} 
Strange things sometimes occur in the other methods
of calculating the Hawking temperature in isotropic coordinates 
since the transformation changes the order of the pole
at the horizon. On the other hand, our formula works nicely. 
The transformation to isotropic coordinates is given
by 
\[
r^\prime = r \left(1 + \frac{m}{2r}\right)^2
\]
so that
\[
g_{00} = - \frac{(2r - m)^2}{(2r+m)^2} 
\]
and
\[
\sqrt{-g} = \frac{(2r-m)(2r+m)}{4r^2} 
\]
where we notice that $\sqrt{-g}$ is zero on the 
horizon. (\ref{generalFormula}) gives
\begin{equation}
\beta = - \frac{i}{2} \oint \frac{(2r+m)^3}{4 r^2 (2r-m)} dr
\end{equation}
and using a partial fraction decomposition 
\begin{equation}
\beta = - \frac{i}{2} \oint \left( 1 + \frac{8m}{2r-m} - \frac{2m}{r} - \frac{m^2}{4 r^2} \right) dr
\end{equation}
and circling the pole at $r=m/2$ (which is the zero of $g_{00}$) 
we get
\begin{equation}
\beta = 8 \pi m
\end{equation}
as expected.

{\bf Rindler space:} 
The form of the Rindler metric 
for an accelerated observer in Minkowski space 
with $g_{00} = -(1 + 2 a r)$ presents no problem. 
On the other hand, the form of the Rindler metric
with $g_{00} = -\alpha^2 r^2$ has
$\sqrt{-g} = \alpha r$. Here, $r_H = 0$ and so
the determinant vanishes on the Horizon. Furthermore,
the usual formula (\ref{generalFormula}) can not
work as it stands since $dr_H/dM$ is not defined. 
However, if we set $dr_H/dM \equiv 1$ we get
\begin{equation}
\beta = - i \oint \frac{\alpha r}{\alpha^2 r^2} dr
= \frac{2 \pi}{\alpha}
\end{equation}
which is the correct temperature. However, even though
it gives the correct answer, this is a problematic case
because our setting $dr_H/dM = 1$ is {\it ad hoc} and 
should be reasoned out more fully. Since $r_H = 0$ and
$M = 0$ for Rindler space one may be able to view this
as a limiting process whereby $r_H \rightarrow 0$ as
$M \rightarrow 0$. 

{\bf Reissner-Nordstr{\" o}m black hole:} 
The metric for the charged black hole has
$g_{00} = -(1 - \frac{2m}{r} + \frac{q^2}{r^2}) = \frac{(r-r_+)(r-r_-)}{r^2}$ 
where $r_\pm = m \pm \sqrt{m^2 - q^2}$.
Integrating around the outer horizon gives
\begin{equation}
\begin{split}
\beta_+ &= - 2 i \oint \frac{z^2}{(r-r_+)(r-r_-)} dr
= 4 \pi \frac{r_+^2}{r_+ - r_-} \\
&= 2 \pi \frac{2m(m+ \sqrt{m^2-q^2}) - q^2}{\sqrt{m^2-q^2}}
\end{split}
\end{equation}
As a side comment, notice that if you integrate around both horizons, 
which leads to summing the residues, the result would
be $\beta_+ + \beta_- = 8 \pi m$. 
We also note that the temperature vanishes in the extremal
$|q| = m$ limit.

{\bf Kerr black hole:} 
The rotating black hole has
$g_{00} = -\frac{\Delta - a^2 \sin^2 \theta}{\Sigma}$ 
where 
$\Delta = (r - r_+)(r-r_-)$ and 
$\Sigma = r^2 + a^2 \cos^2 \theta$. Choose our path at $\theta = 0$
leaving
\begin{equation}
\begin{split}
\beta_+ &= \frac{4\pi m(m+ \sqrt{m^2-a^2})}{\sqrt{m^2-a^2}} \\
\beta_- &= \frac{4\pi m(- m+ \sqrt{m^2-a^2})}{\sqrt{m^2-a^2}} 
\end{split}
\end{equation}
and notice that $\beta_+ + \beta_- = 8 \pi m$ in the same
way as the Reissner-Nordstr{\" o}m black hole and again
the temperature vanishes in the extremal $|a| = m$ limit. 

\section{Conclusion}

We have reviewed the quasi-classical `tunneling' computation
of Hawking radiation, showing that it leads to a simple 
expression (\ref{formula2}) for the Hawking temperature in terms 
of a closed path in complex $r$ space around the horizon.
This can be written in terms of the integral of a 1-form around
the path given by (\ref{form3}) if the time coordinate is transformed 
as in (\ref{timecoord}). 
We have discussed these results in the context of both
black hole thermodynamics and the Euclidean time formalism.

\section{Acknowledgements}

I would like to thank the organizers of the Quarks 2008, 15th 
International Seminar on High Energy Physics in Sergiev Posad, Russia,
where this work was presented. I also gratefully acknowledge conversations
with Nicholas Manton, Dmitri Diakonov, Douglas Singleton and Emil Akhmedov.


\begin{thebibliography}{20}
%
\bibitem{singleton2008} E.T. Akhmedov, Terry Pilling and D. Singleton, arXiv:0805.2635, to appear in Int. J. Mod. Phys. D; 
V. Akhmedova, Terry Pilling, Andrea de Gill and D. Singleton, Phys. Lett. {\bf B666} 269 (2008); 
V. Akhmedova, Terry Pilling, Andrea de Gill and D. Singleton, arxiv:0808.3413 (2008); 
\bibitem{padmanabhan1999} K. Srinivasan and T. Padmanabhan, Phys. Rev. D {\bf 60}, 24007 (1999);
\bibitem{parikh2000} M.K. Parikh and F. Wilczek, Phys. Rev. Lett. {\bf 85}, 5042 (2000);
\bibitem{landau1958} L.D. Landau and E.M. Lifshitz, Quantum Mechanics, Pergamon Press, 1958.
\bibitem{akhmedov2007} E.T. Akhmedov, V. Akhmedova and D. Singleton, 
Phys. Lett. {\bf B642} 124 (2006); 
E.T. Akhmedov, V. Akhmedova, D. Singleton, and Terry Pilling, 
Int. J. Mod. Phys. A {\bf 22} 1705 (2007).
\bibitem{chowdhury2008} B. D. Chowdhury, Pramana {\bf 70}, 593 (2008).
\bibitem{pilling2008} Terry Pilling, Phys. Lett. {\bf B660}, 402 (2008).
\bibitem{zhang2008} Baocheng Zhang, Qing-yu Cai, and Ming-sheng Zhao, Phys. Lett. {\bf B665}, 260 (2008).
\bibitem{hartle1976} J.B. Hartle and S.W.Hawking, Phys. Rev. D {\bf 13}, 2188 (1976).
\bibitem{gibbons1977} G.W. Gibbons and S.W.Hawking, Phys. Rev. D {\bf 15}, 2752 (1977).
%
\end{thebibliography}
\end{document}